\documentclass{JINST}

\newcommand{\degree}{^{\circ}}

\title{The effect of the displacement damage on the Charge Collection Efficiency in Silicon Drift Detectors for the LOFT satellite}

\author{E.~Del Monte$^{a,b}$\thanks{Corresponding
author.}, Y.~Evangelista$^{a,b}$, E.~Bozzo$^c$, F.~Cadoux$^d$, A.~Rachevski$^e$, G.~Zampa$^e$, N.~Zampa$^e$, M.~Feroci$^{a,b}$, M.~Pohl$^d$, A.~Vacchi$^{e,f}$\\
\llap{$^a$}INAF - Istituto di Astrofisica e Planetologia Spaziali, \\
  Via Fosso del Cavaliere 100, I-00133 Roma, Italy\\
\llap{$^b$}INFN - Istituto Nazionale di Fisica Nucleare, Sezione di Roma Tor Vergata,\\
  Via della Ricerca Scientifica 1, I-00133  Roma, Italy\\
\llap{$^c$}ISDC - Data Centre for Astrophysics, Universit\'{e} de Gen\`{e}ve,\\
  Chemin d'Ecogia 16, CH-1290 Versoix, Switzerland\\
\llap{$^d$}DPNC - D\'{e}partement de Physique Nucl\'{e}aire et Corpusculaire, Universit\'{e} de Gen\`{e}ve, \\
  Quai Ernest-Ansermet 24, CH-1211 Gen\`{e}ve, Switzerland\\
\llap{$^e$}INFN - Istituto Nazionale di Fisica Nucleare, Sezione di Trieste,\\
  Padriciano 99, I-34149 Trieste, Italy\\
\llap{$^f$}Mathematics and Informatics Departement, Udine University\\
  Via delle Scienze 206, I-33100 Udine, Italy\\
  E-mail: \email{ettore.delmonte@iaps.inaf.it}}

\abstract{The technology of Silicon Drift Detectors (SDDs) has been selected for the two instruments aboard the Large Observatory For X-ray Timing (LOFT) space mission. LOFT underwent a three year long assessment phase as candidate for the M3 launch opportunity within the ``Cosmic Vision 2015 -- 2025'' long-term science plan of the European Space Agency. During the LOFT assessment phase, we studied the displacement damage produced in the SDDs by the protons trapped in the Earth's magnetosphere. In a previous paper we discussed the effects of the Non Ionising Energy Losses from protons on the SDD leakage current. In this paper we report the measurement of the variation of Charge Collection Efficiency produced by displacement damage caused by protons and the comparison with the expected damage in orbit.}


\begin{document}

\section{Introduction}
\label{sec:introduction}

Silicon Drift Detectors (SDDs) are semiconductor devices introduced in the 1980s as position sensing detectors for high energy charged particles and X-ray spectroscopy \cite{Gatti_Rehak_1984,Gatti_Rehak_2005}. The  Inner Tracking System of the ALICE detector at the CERN Large Hadron Collider includes 260 linear and large area SDDs \cite{ALICE_LHC,Vacchi_et_al_1991}, for a total active area of 1.4 m$^2$, employed to measure the trajectories of charged particles. Large area linear SDDs have been selected for the two instruments aboard the satellite-borne Large Observatory For x-ray Timing (LOFT\footnote{In this paper we refer to the LOFT configuration for the M3 launch opportunity.} \cite{LOFT_2012}): the Large Area Detector and the Wide Field Monitor. The Large Area Detector (LAD \cite{LAD_SPIE_2012}) is a collimated instrument for X-ray timing and, in the M3 design, it had a geometric area of $\sim 15 \; \mathrm{m}^2$, covered by 2016 SDDs. This design led to an effective area for X-ray photons of $\sim 10 \; \mathrm{m}^2$ at 8 keV, a field of view of $\sim 1 \degree$, and a spectral resolution of $\sim 240$ eV Full-Width at Half Maximum (FWHM) at 6 keV (see ref. \cite{LAD_SPIE_2012} for more details). The Wide Field Monitor (WFM \cite{WFM_SPIE_2012}) is a coded aperture imager for X-rays and, in the M3 design, it is composed of five independent units, each one made of two co-aligned cameras, with an instantaneous field of view of $\sim 4.1$ sr. With such a design, the WFM can simultaneously observe about one third of the sky with an angular resolution of $\sim 5$ arcmin and a location accuracy of $\sim 1$ arcmin for point sources.

The LOFT SDDs are subdivided into two symmetrical half-detectors, each one with a set of collecting anodes implanted at the edge \cite{Rachevski_et_al_2014}. The charge cloud produced after the interaction of an X-ray photon is transported toward the anodes using a constant electric field, sustained by a negative voltage applied to a series of cathodes implanted on the top and bottom faces of the detector, whose value progressively decreases approaching the anodes at $\sim 0$ V. With this  method, the signals produced by the interaction of photons in a sensitive area of tens of cm$^2$ can be collected using anodes with a surface of tenths of mm$^2$ each and a capacitance of the order of a hundred fF. The working principle, the design methodology and the performance of the SDDs prototypes produced during the LOFT M3 assessment phase are described in Ref. \cite{Rachevski_et_al_2014}.

Silicon detectors are known to be sensitive to the displacement damage produced by charged and neutral particles, i.e. the creation of crystal defects with energy states within the semiconductor band-gap. Defects located in the depleted region, with energy levels near the middle of the gap, behave as generation centers that contribute an additional component to the detector leakage current \cite{Segneri_et_al_2009}. In addition, damage centers in the lattice having energy levels close to the valence and conduction bands are able to trap the charge carriers, thus reducing the device Charge Collection Efficiency (CCE) \cite{Kramberger_et_al_2002}.

X-ray timing studies are based on the measurement of the flux of astrophysical sources, in search of periodicities, quasi-periodic oscillations, and other similar features. In order to maximise the accuracy of the flux measurement, a high count rate is required. For this reason, any mechanism affecting the detection and collection of photons, such as the detector quantum efficiency, CCE or dead time, needs to be carefully studied and accounted for.

We have described in Ref. \cite{Del_Monte_et_al_2014} the radiation environment expected for the LOFT orbit. Although the baseline orbit identified during the assessment phase had an altitude of 550 km and an inclination of $0 \degree$, in this work we take into account as a worst case the less favourable orbit at 600 km altitude and $5 \degree$ inclination. The nominal duration of the LOFT mission studied for M3 was 4.25 years, including three months of commissioning phase.

As discussed in Ref. \cite{Del_Monte_et_al_2014}, the LOFT radiation environment is dominated by protons, especially those trapped in the South Atlantic Anomaly. For this reason we restrict the study presented here to the proton component, neglecting the other types of particles. We estimated the proton flux using the AP8 model in Minimum Solar Conditions (AP8-MIN) included in the SPENVIS web-based software\footnote{\texttt{http://www.spenvis.oma.be/}} \cite{SPENVIS_2002,SPENVIS_2009,SPENVIS_2010}. The flux in AP8-MIN is higher than in AP8-MAX (Solar Maximum Conditions), thus our choice represents a conservative case. For example, for an orbit at 600 km altitude and $5 \degree$ inclination, the integral flux of AP8-MIN is $\sim 7$ times higher than AP8-MAX at 1 MeV, and $\sim 6$ times higher at 10 MeV.

The contribution of trapped protons may be evaluated in SPENVIS as the equivalent fluence at a single energy, selected by the user and transported through the shielding thickness (assumed as a spherically symmetric aluminum layer). In our estimation we adopt for the protons the default energy value of 10 MeV and for the shielding materials around the SDDs a thickness equivalent to 3.3 mm of aluminum, as estimated for the LAD in Ref. \cite{Del_Monte_et_al_2014,Campana_et_al_2013}.

The evaluation of the expected proton fluence at the LOFT orbit and the measurement of the increase in leakage current after the irradiation of two SDD prototypes with a proton beam was reported in Ref. \cite{Del_Monte_et_al_2014}. As a follow up of that work, we describe here the measurement of the variation of the CCE of an SDD after the proton irradiation and we compare the results with the predictions. In section \ref{sec:CCE_formulas} we summarise the formulae describing the CCE of semiconductor detectors and the variation produced by the displacement damage. In section \ref{sec:measurements} we describe the method and the experimental set-up used to measure the CCE. The details of the irradiation are given in section \ref{sec:Irradiation}. In section \ref{sec:Results} we discuss the results and finally in section \ref{sec:conclusions} we summarise our findings and draw our conclusions.


\section{The Charge Collection Efficiency in silicon drift detectors}
\label{sec:CCE_formulas}

The signal of a semiconductor detector is a charge cloud produced after the interaction of a photon or a charged particle. The detector CCE is influenced by any phenomenon able to reduce the number of charge carriers in the cloud, between the production and the collection sites. For example, impurities and defects in the semiconductor lattice, including those produced by the displacement damage, can act as trapping centers. These centers remove free charge carriers from the signal, thus reducing the detector CCE. In this section we summarise the equations that are usually considered to describe the variation of the CCE with the displacement damage (see Ref. \cite{Kramberger_et_al_2002,Kramberger_et_al_2007,Spieler_2005}).

If the thermal velocity of a charge is higher than the drift velocity, the integrated path length of the charge, and thus the number of collisions, is proportional to the drift time and thermal velocity \cite{Kramberger_et_al_2002}. Under this assumption, during the drift the value of the charge cloud $q$ as a function of time $t$ is given by \cite{Kramberger_et_al_2002}

\begin{equation}\label{eq:charge_vs_time}
    q(t) = q(0) \; e^{-t / \tau} 	
\end{equation}

\noindent where $\tau$ is a time constant related to the charge lifetime for all the trapping processes. For a collection time $t_c$,

\begin{equation}\label{eq:CCE_vs_time}
    CCE = \frac{q(t_c)}{q(0)} = e^{-t_c / \tau}
\end{equation}						

\noindent The collection time depends on the drift velocity of the charge in the semiconductor,

\begin{equation}\label{eq:v_drift}
    v_{drift} = \mu  E
\end{equation}

\noindent where $E$ is the electric field ($360 \; \mathrm{V \; cm^{-1}}$ for the LOFT SDDs) and $\mu$ is the charge mobility. The charge carriers in the SDDs are electrons and their mobility $\mu_e$ changes with the absolute temperature $T$ as \cite{Jacoboni_et_al_1977}

\begin{equation}\label{eq:mobility}
    \mu_e = 1400 \times \left( \frac{300 \; K}{T} \right)^{2.42} \; \mathrm{cm^2 \; V^{-1} \; s^{-1}}
\end{equation}

\noindent Here we verify that, as previously stated and specified in Ref. \cite{Kramberger_et_al_2002,Spieler_2005}, in the temperature range of the LOFT SDDs the thermal velocity is higher than the drift velocity. In a semiconductor the thermal velocity of the electrons is given by

\begin{equation}\label{eq:v_thermal}
    v_{thermal} = \sqrt{\frac{4 \; k_B T}{3 \; m_e^*}}
\end{equation}

\noindent where $k_B$ is the Boltzmann constant and $m_e^* = 0.26 \; m_e$ in silicon, with $m_e$ electron mass ($9.109 \times 10^{-31}$ kg). Typical values of the temperature, electron mobility, thermal velocity, drift velocity and maximum collection time for the LOFT SDDs are listed in Tab. \ref{tab:T_mu_V_drift_t_c}. As shown in this table, the thermal velocity of the carriers is more than one order of magnitude higher than the drift velocity, thus we can apply Eq. (\ref{eq:charge_vs_time}) and Eq. (\ref{eq:CCE_vs_time}).

The maximum collection time is measured for photons impinging exactly at the center of the detector tile, i.e. at the longest drift length (3.5 cm) from the collection anodes. Since the drift length is the same for the SDDs in the LAD and WFM and the upper boundary of the operative temperature range is similar ($-3 \; \degree$C for the WFM and $-10 \; \degree$C for the LAD), the collection time for the two instruments differs only by 6 \% (see Table \ref{tab:T_mu_V_drift_t_c}).

\begin{table}[h!]
\centering

\caption{Typical values of the temperature ($T$), electron mobility ($\mu_e$), thermal velocity ($v_{thermal}$), drift velocity ($v_{drift}$), and maximum collection time ($t_c$) for the LOFT SDDs. In the calculation we assume a drift length of 3.5 cm and an electric field of $360 \; \mathrm{V \; cm^{-1}}$. The upper boundary of the LAD operative temperature is $-10 \; \degree$C \cite{Del_Monte_et_al_2014}, and the measurements of the CCE have been performed at $-38 \; \degree$C (see section 3 below).} 

\begin{tabular}{c c c c c l}
  \hline

  $T$            & $\mu_e$                               & $v_{thermal}$             & $v_{drift}$               & $t_c$              & Comment \\
  $[\degree$C] & $[\mathrm{cm^2 \; V^{-1} \; s^{-1}}]$ & $[\mathrm{cm \; s^{-1}}]$ & $[\mathrm{cm \; s^{-1}}]$ & $[\mathrm{\mu s}]$ & \\
  \hline
  $-38$        & $2.5 \times 10^3$                     & $1.35 \times 10^7$        & $9.1 \times 10^5$         &  3.9               & Measurements in this paper \\
  $-10$        & $1.9 \times 10^3$                     & $1.43 \times 10^7$        & $6.9 \times 10^5$         & 5.1                & LAD operative temperature\\
  $-3$         & $ 1.8 \times 10^3$                     & $ 1.45 \times 10^7$        & $ 6.5 \times 10^5$         & 5.4              & WFM operative temperature\\
  $+20$        & $1.5 \times 10^3$                     & $1.51 \times 10^7$        & $5.3 \times 10^5$         & 6.6                & Room temperature \\
  \hline
\end{tabular}\label{tab:T_mu_V_drift_t_c}
\end{table}

After the irradiation of the detector with a particle fluence $\Phi$, the time constant $\tau$ in Eq. (\ref{eq:charge_vs_time}) and Eq. (\ref{eq:CCE_vs_time}) is such that \cite{Kramberger_et_al_2002}

\begin{equation}\label{eq:tau_fluence}
    \frac{1}{\tau} \sim \beta \; \Phi
\end{equation}

\noindent and is dominated by the contribution from the radiation damage. In particular, from Eq. (\ref{eq:charge_vs_time}) and Eq. (\ref{eq:CCE_vs_time})

\begin{equation}\label{eq:1-CCE}
    1 - CCE \sim \beta \; \Phi \;  t_c
\end{equation}

\noindent The constant $\beta$ is measured in Ref. \cite{Kramberger_et_al_2002} for holes and electrons as charge carriers and for an irradiation with protons, neutrons, and pions. The authors remark that $\beta$ depends on the type of particle used in the irradiation, consequently the Non Ionising Energy Losses (NIEL) scaling hypothesis cannot be applied to the study of the CCE. Conversely, the NIEL hypothesis is valid for example to compute the increase in leakage current produced by the displacement damage \cite{Segneri_et_al_2009}. As specified in Ref. \cite{Kramberger_et_al_2002}, the constant $\beta$ follows a power-law behaviour as a function of temperature $T$,

\begin{equation}\label{eq:beta_vs_temperature}
    \beta (T)  =  \beta (T_0) \; \left( \frac{T}{T_0} \right)^{\alpha}
\end{equation}

\noindent In the LOFT case the charge carriers are electrons and the largest displacement damage is produced by protons \cite{Del_Monte_et_al_2014}. In this case, from Ref. \cite{Kramberger_et_al_2002} we find that in Eq. (\ref{eq:beta_vs_temperature}) $\alpha  = -0.86$ and $\beta = 5.6 \times 10^{-16} \; \mathrm{cm^2 \; ns^{-1}}$ at $T_0 = 263$ K (i.e. $-10 \; \degree$C).

The dependence of the damage annealing on time and temperature is \cite{Kramberger_et_al_2002,Kramberger_et_al_2007}

\begin{equation}\label{eq:beta_vs_time}
    \beta(t) = \beta_0 \; e^{-t / \tau_A} + \beta_{\infty} \; (1 - e^{-t / \tau_A})
\end{equation}

\noindent and the time constant $\tau_A$ follows the Arrhenius relation

\begin{equation}\label{eq:tau_annealing}
    \tau_A = \tau_0 \; e^{E / k_B T}
\end{equation}

\noindent where, for electrons as charge carriers, $\tau_0 = 3.88 \times 10^{-14}$ min and $E = 1.06$ eV \cite{Kramberger_et_al_2007}.

For the operational temperature of the SDDs in the LAD, $\tau_A \geq 14.8$ years, longer than the mission duration (4.25 years). Consequently we can neglect the annealing when estimating the variation of the CCE in LOFT.

\section{Measurement of the CCE}
\label{sec:measurements}

\subsection{Measurement technique}

The total charge of the electron cloud produced in an SDD after the interaction of an X-ray is proportional to the photon energy. On average the energy to produce an electron-hole pair is 3.6 eV. The charge cloud is then drifted toward the anodes by means of the electric field. Any mechanism affecting the device CCE reduces the amount of charge collected at the anodes, and thus the amplitude of the signal provided in output by the detector.

In order to measure the variation of the CCE produced by the displacement damage, we compared the spectrum of an X-ray line from a source at the same location on the detector, before and after an irradiation with protons. Any fractional variation of the CCE gives an equal fractional variation of the peak position (in ADC channels) of the X-ray line in the detected spectrum.

\subsection{Characteristics of the detector under test}

We irradiated an SDD prototype of the XDXL-1 production batch by Fondazione Bruno Kessler (FBK of Trento in Italy) which was designed by INFN-Trieste. The SDD had a thickness of 450 $\mu$m, a drift length of 3.5 cm, and a geometric area of 5.52 cm (in the anode direction) $\times$ 7.25 cm (in the drift direction). The anode pitch was different in the two detector halves: 835 $\mu$m in the ``LAD half'' and 294 $\mu$m in the ``WFM half''. More details about the design and characteristics of the different prototypes of SDDs for LOFT are reported in Ref. \cite{Rachevski_et_al_2014}.

\begin{figure}[h!] \centering
\includegraphics[width=12 cm]{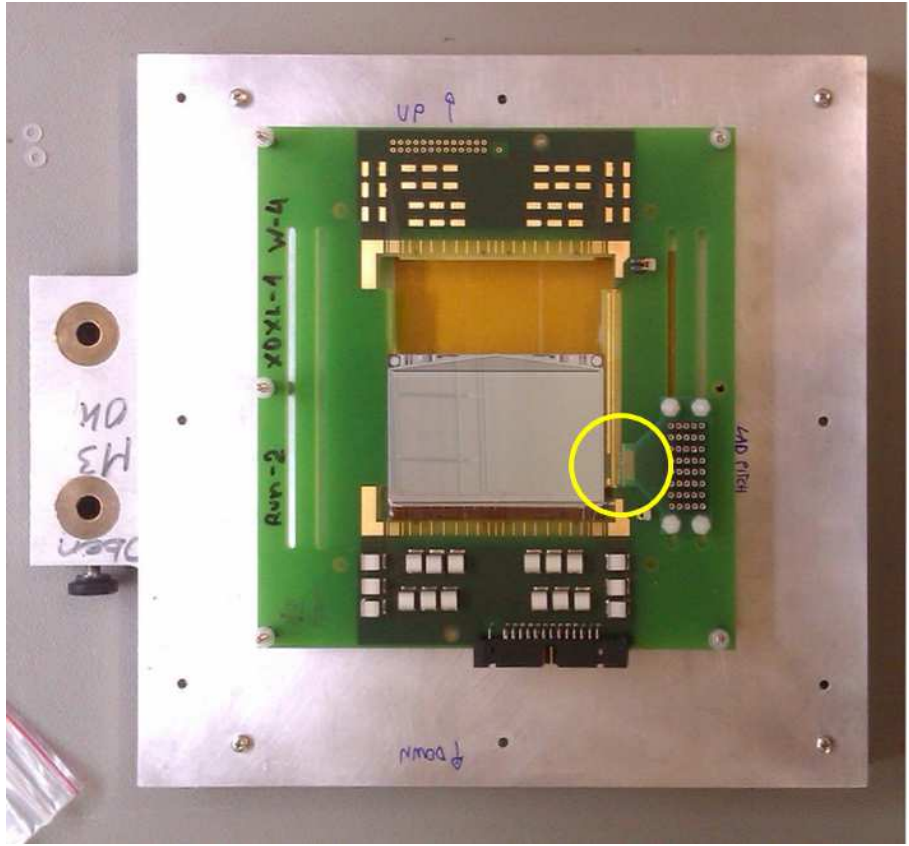}
\caption{Picture of the XDXL-1 detector mounted on the PCB and the aluminum support for the irradiation. The yellow circle highlights the position of the eight anodes connected to the discrete FEE. The SDD has a thickness of 450 $\mu$m, a drift length of 3.5 cm, and a geometric area of 5.52 cm (in the anode direction) $\times$ 7.25 cm (in the drift direction).} \label{fig:SDD_PCB}
\end{figure}

A set of eight contiguous channels in the LAD half, highlighted by the yellow circle in the picture in Fig. \ref{fig:SDD_PCB}, was connected to a Front-End Electronics (FEE) based on discrete components. Each anode was individually connected to an analogue chain composed of a low gate capacitance SF-51 JFET (with capacitance of 0.4 pF) used as the input transistor of an Amptek A250F-NF charge sensitive amplifier. The feedback capacitor of 50 fF and a reset transistor were both integrated on the JFET die, allowing to reduce the input stray capacitance for a better noise performance \cite{Zampa_et_al_2011,Campana_et_al_2011}. The Test Equipment (TE) used in the measurements allowed to trigger only on one anode, for which we set a threshold of 200 ADC channels (corresponding to $\sim 3.5$ keV).

\subsection{Experimental set-up}

In the characterisation we used the manganese $K_{\alpha}$ line at 5.9 keV, emitted by an $^{55}$Fe source with activity of $\sim 1.2$ mCi. The source also produces a fainter $K_{\beta}$ line at 6.5 keV, with a branching ratio $\sim 8.5$ times smaller than the $K_{\alpha}$.

The proton fluence during the irradiation was expected to produce a small decrease of the CCE but a significant increase (2.4 nA/anode at 20 $\degree$C) of the bulk leakage current. At the same temperature, the noise introduced by this additional leakage current would give an FWHM above 1 keV at the energy of the $K_{\alpha}$  line even with the minimum peaking time selectable in the TE. The value of the CCE is almost insensitive to temperature, as shown in Eq. (\ref{eq:mobility}) and Eq. (\ref{eq:beta_vs_temperature}). The leakage current leading to an increase of FWHM is characterised by an exponential dependence on temperature \cite{Spieler_2005}. For this reason, we characterised the SDD at about $-38 \; \degree$C inside a thermal chamber. The detector temperature was measured using an AD590 thermometer, placed inside the detector box. By working at low temperature, we mitigated the increase of the FWHM while leaving almost unaffected the variation of the CCE, thus enhancing the sensitivity of our spectral measurements.

\begin{figure}[h!] \centering
\includegraphics[width=12 cm]{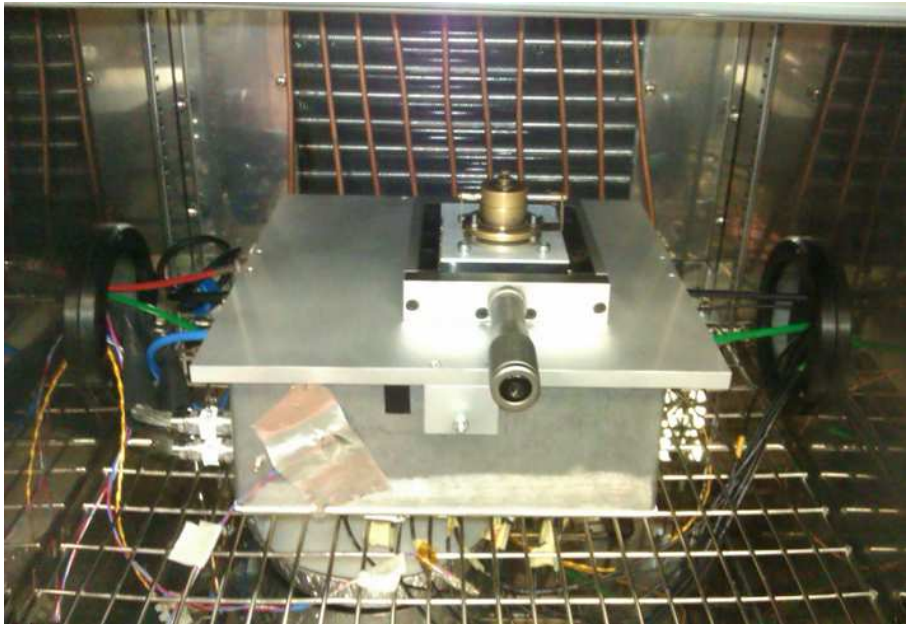}
\caption{Picture of the experimental set-up for the SDD characterisation: the box containing the XDXL1 detector, the translation stage and the brass container with the $^{55}$Fe source.} \label{fig:SDD_thermal_chamber}
\end{figure}

Given the difficulty to obtain the absolute source position at a level of fractions of millimeter with the set-up available inside the thermal chamber, we decided to use the relative distance instead. We compared the spectra acquired at three different distances along the drift channel from a common origin ($x_0  \simeq 3$ mm from the anodes): 0 mm (near the anodes), 15 mm (mid-distance) and 30 mm (at the end of the drift channel). The source was positioned using a micrometric translation stage. A picture of the detector box in the thermal chamber with the translation stage and the brass container of the $^{55}$Fe source is shown in Fig. \ref{fig:SDD_thermal_chamber}. With this method, we kept the relative distance between the three positions of the source fixed, thus relaxing the constraints on the absolute location.

We collimated the source with a slit (1 mm $\times$ 10 mm) and a diaphragm (400 $\mu$m aperture) and we found that the footprint of the beam on the detector was 1.5 mm $\times$ 0.75 mm, i.e. still on the same anode of the LAD half. With this set-up, we obtained a count rate of $\sim 6 \; \mathrm{cts \; s^{-1}}$. Due to diffusion, the size of the electron cloud produced by the interaction of an X-ray photon increases during the drift toward the anodes. The expected cloud size for the maximum drift length of 3.5 cm was $\sim 1$ mm \cite{Zampa_et_al_2011,Campana_et_al_2011}. Since the anode pitch of the SDD under test was 835 $\mu$m, the charge spread on 1 -- 3 anodes. The acquisition system was triggered by a single fixed channel, whose signal was then summed to that of the two neighbouring ones (the closest on the left and right) to determine the total amount of charge deposited on the detector. Due to the fixed trigger configuration and the large collimator slit, we selected for the analysis only the events when no more than 25 \% of the signal was collected by the side channels, in order to ensure complete charge collection by removing events triggering on the tails of the signal charge distribution. We applied the same method for all the three locations of the source along the drift channel specified above.

For each location, we accumulated one spectrum before and one after the irradiation. The exposure was 3600 s for all the measurements. We subtracted from the signal the common mode noise (estimated by averaging the amplitude of the other five channels in the set) and the pedestals of the ADC (independently estimated on each read-out channel with a specific measurement using the same TE). After the subtraction, we fitted each spectrum with two Gaussians, whose distance between the peaks and ratio of the peak count rates was fixed based on the properties of the manganese $K_{\alpha}$ (5.9 keV) and $K_{\beta}$ (6.5 keV) fluorescence lines.

\subsection{Characterisation of the SDD before the irradiation}

\begin{figure}[th!] \centering
\includegraphics[width=12 cm]{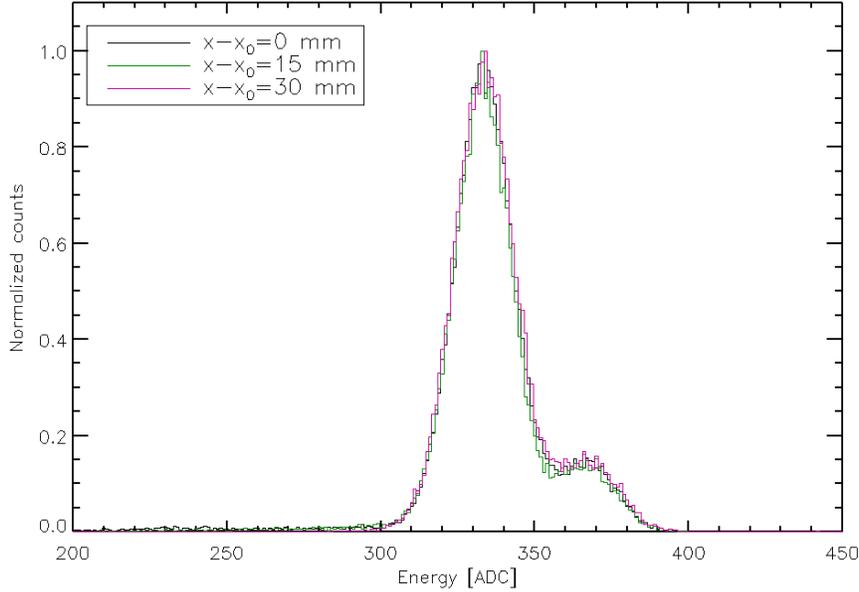}
\caption{Superposition of the spectra of the $^{55}$Fe source placed at the three values of the distance from the common origin ($x_0$): 0 mm (near the anodes, black line), 15 mm (mid-distance, green line) and 30 mm (at the end of the drift channel, magenta line) before the irradiation. The temperature during the measurement was $-38 \; \degree$C. The histograms are normalised to the maximum counts per bin.} \label{fig:spectrum_pre}
\end{figure}

We show in Fig. \ref{fig:spectrum_pre} a superposition of the spectra accumulated near the anodes, at mid-distance and at the end of the drift channel, before the irradiation. The spectra are shown after the subtraction of the pedestals of the ADC. The results of the fits are listed in Tab. \ref{tab:fit_before_irradiation}. By moving the source of 30 mm we found that the fitted position of the peak in the spectrum changed by $\sim 0.2$ ADC channels (corresponding to $\sim 3$ eV), which is negligible considering the energy resolution of the measurement. For all purposes, the collection efficiency before the irradiation can be considered as 100 \%.

As expected, the FWHM does not significantly change in the three locations along the drift channel. Due to the charge diffusion during the drift, in the experimental conditions the width of the charge cloud at the anodes (at $\pm 3 \sigma$) ranges between $\sim 350 \; \mu$m and $\sim 1150 \; \mu$m respectively for photon interactions close to the collecting electrodes and at the maximum drift distance. For this reason, the fraction of the events fulfilling the selection criterion described above decreases with the distance from the anodes, explaining the variation of the total count rate with distance, shown in Tab. \ref{tab:fit_before_irradiation}. We adjusted the signal shaping time to make negligible the ballistic deficit that otherwise could have an impact on the energy measurement at different drift times.

\begin{table}[h!]
\centering
\caption{Parameters of the fit of the $^{55}$Fe $K_{\alpha}$ X-ray line before the irradiation. The FWHM is measured on the $K_{\alpha}$ line, the total count rate on $K_{\alpha}$ and $K_{\beta}$ lines.}
\begin{tabular}{c c c}
  \hline
  Distance from the anodes & FWHM   & Total count rate \\
  $[$mm$]$                 & [eV]   & [$\mathrm{counts \; s^{-1}}$] \\
  \hline
  0                        & 394    & 15.8\\
  15                       & 395    & 12.0 \\
  30                       & 405    & 10.9 \\
  \hline
\end{tabular}\label{tab:fit_before_irradiation}
\end{table}

\subsection{Characterisation of the SDD after the irradiation}

\begin{figure}[bh!] \centering
\includegraphics[width=12 cm]{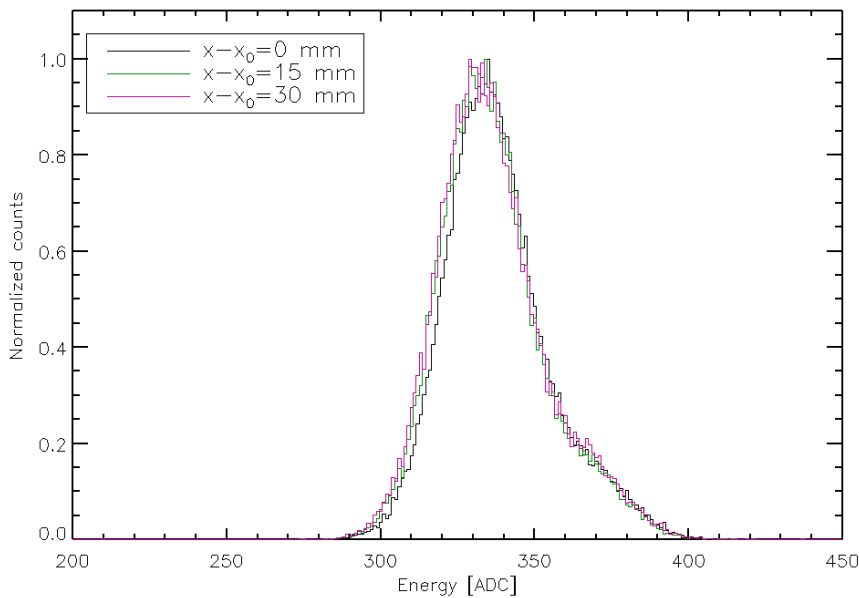}
\caption{Superposition of the spectra of the $^{55}$Fe source placed at the three values of the distance from the common origin ($x_0$): 0 mm (near the anodes, black line, $-38 \; \degree$C), 15 mm (mid-distance, green line, $-38 \; \degree$C) and 30 mm (at the end of the drift channel, magenta line, $-37 \; \degree$C). The histograms are normalised to the maximum counts per bin. The increase of the FWHM in the spectra after the irradiation is due to the additional leakage current, only partially compensated by the low temperature.} \label{fig:spectrum_post}
\end{figure}

After the irradiation, we repeated the characterisation of the SDD described above. The superposition of the three spectra (at the end of the drift channel, at mid-distance and near the anodes) is shown in Fig. \ref{fig:spectrum_post}. The spectra are shown after the subtraction of the ADC pedestals. The results of the fit are listed in Tab. \ref{tab:fit_after_irradiation}.

\begin{table}[h!]
\centering
\caption{Parameters of the fit of the $^{55}$Fe $K_{\alpha}$ X-ray line after the irradiation. The FWHM is measured on the $K_{\alpha}$ line, while the total count rate on $K_{\alpha}$ and $K_{\beta}$ lines.}
\begin{tabular}{c c c}
  \hline
  Distance from the anodes & FWHM   & Total count rate \\
  $[$mm$]$                 & [eV]   & [$\mathrm{counts \; s^{-1}}$] \\
  \hline
  0                        & 536    & 15.9 \\
  15                       & 548    & 12.4 \\
  30                       & 555    & 10.7 \\
  \hline
\end{tabular}\label{tab:fit_after_irradiation}
\end{table}

The increase of the FWHM in the spectra after the irradiation is due to the additional leakage current, only partially compensated by the measurement at low temperature. Similarly to the measurements before the irradiation, the FWHM does not significantly change in the three locations along the drift channel.

\section{Irradiation of the SDD}
\label{sec:Irradiation}

We irradiated the SDD at the low energy site of the Proton Irradiation Facility (PIF) in the accelerator of the Paul Scherrer Institute\footnote{\texttt{http://pif.web.psi.ch/}} (PSI) in Villigen (Switzerland). The facility provides a proton beam with a maximum flux of $\sim 5 \times 10^8 \; \mathrm{cm^{-2} \; s^{-1}}$ and  energy ranging in quasi-continuous mode between $\sim 6$ and $\sim 63$ MeV using degrader foils \cite{Hajdas_et_al_1996}.

During the irradiation, the proton spectrum was centered at 11.2 MeV. This is the closest energy value available at the facility to match the energy that we used for the estimation with SPENVIS (10 MeV). The beam spectrum had a gaussian shape, with an FWHM of $\sim 6$ MeV. The irradiation was performed in air and at room temperature (about 25 $\degree$C).

\begin{figure}[h!]\centering
\includegraphics[width=12 cm]{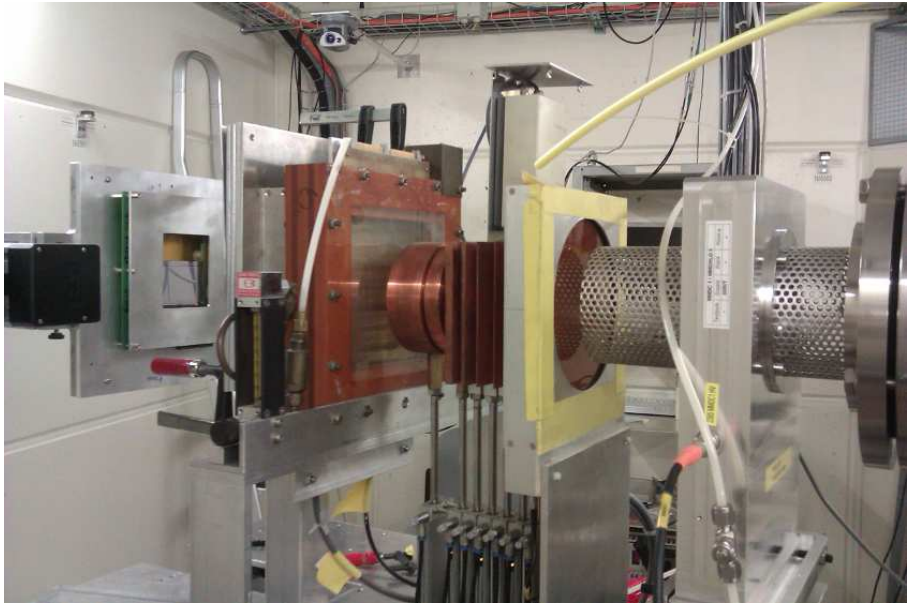}
\caption{Picture of the experimental set-up during the irradiation. On the left the movable table and the sample holder are shown. The direction of the proton beam in the picture is from right to left.}\label{fig:PIF_PSI_set-up}
\end{figure}

The facility was equipped with a sample holder fixed on a movable table, which could be displaced in horizontal, vertical and longitudinal directions. We specifically designed and produced a mechanical structure to fix the device under test to the sample holder (see Fig. \ref{fig:PIF_PSI_set-up}). A laser cross-hair system was used as a reference to align the movable table with the geometrical center of the proton beam. In order to irradiate only the SDD and not the other electronic components, the PCB was shielded with a $\sim 3$ mm thick aluminum layer, able to stop all the protons in the beam, with a hole in the center to expose only the SDD (see the picture in Fig. \ref{fig:PIF_PSI_set-up}).

\begin{figure}[h!] \centering
\includegraphics[angle=90, width=12 cm]{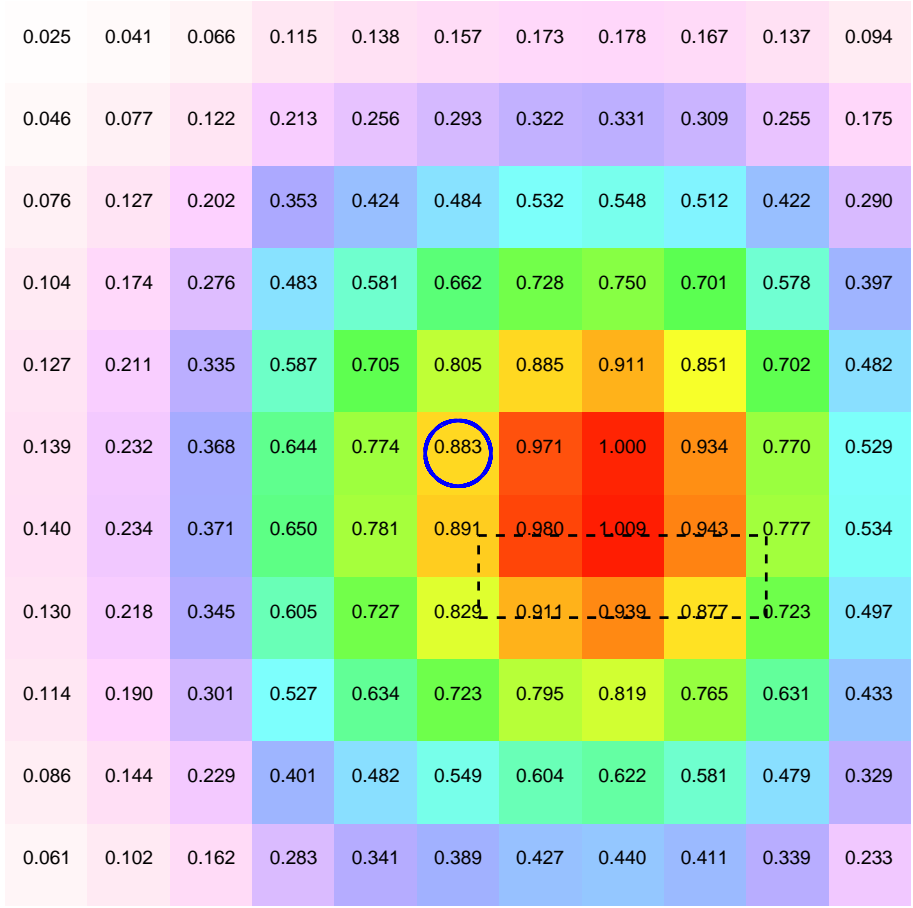}
\caption{Map of the reconstructed beam uniformity, in 1 cm steps, during the irradiation. The center of the beam is indicated by the blue circle. The horizontal $x$ axis (left to right) is along the direction of the charge drift. The vertical $y$ axis (up to down) is along the direction of the readout anodes. The dashed black box approximately defines the contour (1 cm $\times$ 3.5 cm) of the anodes connected to the FEE, where the average fraction of the beam flux was 88.8 \%.} \label{fig:beam_map}

\end{figure}

The beam was normally incident on the detector during the irradiation and was defocussed in order to cover the whole surface. The shape and uniformity of the beam were derived before the irradiation from measurements of the proton flux with a motorised and remotely controlled scintillation detector, at steps of 1 cm along both the horizontal ($x$ axis) and vertical ($y$ axis) directions, for a length of 11 cm on each axis. We computed for each step the fraction of the flux with respect to the maximum. The beam profile had a gaussian shape and we independently applied gaussian fits to the fraction values along the $x$ and $y$ axis in order to derive the intensity of the beam profile. In addition, we reconstructed the map of the beam fraction on the surface of 11 cm $\times$ 11 cm by multiplying the fractions along $x$ and $y$, as shown in Fig. \ref{fig:beam_map}. During this calibration, we found that the maximum of the flux was in $x \simeq 2$ cm and $y \simeq 0$ cm and that the flux had also the highest uniformity in the region around the maximum (see Fig. \ref{fig:beam_map}). We thus put the geometric center of the LAD half at $x \simeq 2$ cm and $y \simeq 0$ cm during the irradiation, in order to exploit the region with the maximum flux and the highest uniformity. The average fraction of the beam intensity on the surface of the eight anodes connected to the FEE was 88.8 \% (highlighted in Fig. \ref{fig:beam_map}).

We selected for the irradiation a fluence corresponding to ten times the value at 600 km and $5 \degree$ for a mission duration of 4.25 years \cite{Del_Monte_et_al_2014}, more than one order of magnitude higher than the value predicted for the nominal LOFT orbit (including the margin of a factor of 20 applied to derive the operational temperature). This fluence was expected to produce a measurable variation of the CCE with the experimental set-up available in the laboratory and to allow us to compare the CCE with the prediction from Eq. (\ref{eq:1-CCE}). On average, the fluence on the surface of the eight anodes connected to the FEE was $7.9 \times 10^8 \; \mathrm{cm^{-2}}$ and was provided in a single exposure, with a flux of $1.8 \times 10^6 \; \mathrm{cm^{-2} \; s^{-1}}$.

As reported in Ref. \cite{Del_Monte_et_al_2014}, two sets of wire chambers, one upstream and the other downstream with respect to the degrader foils, were used to monitor the beam flux during the irradiation. The uncertainty on the measure of the fluence was due to the time variability of the beam. We assumed as uncertainty on the fluence the ratio between the standard deviation and the average value of the count rate, measured every second by the downstream wire chamber. The resulting uncertainty was 2.4 \% of the total fluence.

\section{Results and discussion}
\label{sec:Results}

\subsection{Variation of the CCE}

\begin{figure}[h!] \centering
\includegraphics[width=12 cm]{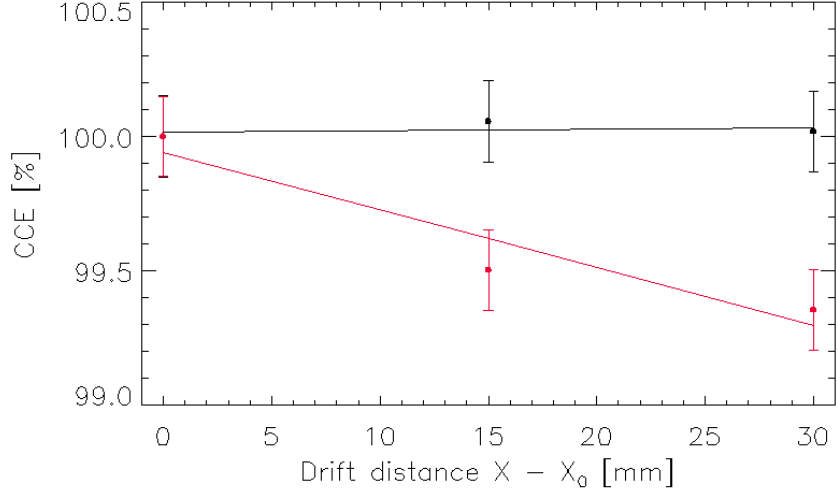}
\caption{CCE derived from the position of the reconstructed peak of the manganese $K_{\alpha}$ line as a function of distance from the anodes before (black line) and after (red line) the irradiation.} \label{fig:CCE_distance}
\end{figure}

We show in Fig. \ref{fig:CCE_distance} the CCE, derived from the position of the peak of the manganese $K_{\alpha}$ line reconstructed from the fit of the spectra. The figure contains the spectra acquired near the anodes, at the end of the drift channel and at mid-distance, before and after the irradiation. As shown in the figure, the measured decrease of the CCE at a distance of 30 mm from the reference position near the anodes is $(0.65 \pm  0.15)$ \%. The uncertainty on the CCE is derived from the uncertainty on the peak position, from the fitting procedure described in section \ref{sec:measurements}.

From Eq. (\ref{eq:1-CCE}), with the provided fluence ($7.9 \times 10^8 \; \mathrm{cm^{-2}}$ on average on the eight anodes connected to the FEE) and $\beta$ computed at $-37 \; \degree$C from Eq. (\ref{eq:beta_vs_temperature}), the expected reduction of the CCE is 0.8 \%, compatible with the measurements.

\subsection{Variation of the bulk leakage current}

The measurement of the variation of the bulk leakage current after the irradiation is reported in Ref. \cite{Del_Monte_et_al_2014}. We recall here, for the sake of completeness, that the increase of leakage current, measured $\sim 6.3$ days after the irradiation following the method described in Ref. \cite{Del_Monte_et_al_2014}, was 2.4 nA/anode at 20 $\degree$C. This value is remarkably consistent with the expected increase of 2.4 nA/anode, estimated with the method in Ref. \cite{Segneri_et_al_2009,Moll_et_al_2002} and the following assumptions: (1) the average fraction of the beam intensity on the surface of the eight anodes connected to the FEE was 88.8 \%; (2) the residual damage after $\sim 6.3$ days of annealing at room temperature was 44 \% of the initial one. All the details are described in Ref. \cite{Del_Monte_et_al_2014}.

\subsection{Discussion}

The variation of the CCE produced by the Total Ionising Dose (TID) in silicon detectors is discussed for example in Ref. \cite{Fourches_2011,Ratti_et_al_2014}. The authors show that the decrease of CCE is negligible for a TID of tens \cite{Fourches_2011} or even hundreds \cite{Ratti_et_al_2014} of krad. Consequently, the TID of $\sim 400$ rad(Si) provided to the SDD during our irradiation at PIF is not expected to produce significant effects on the CCE.

The characterisation of the SDD was concluded 4.2 days after the end of the irradiation. Assuming a storage temperature of 25 $\degree$C, the time constant for the annealing from Eq. (\ref{eq:tau_annealing}) is 22.8 days. The interval of time between the irradiation and the measurement is thus $\sim 1/5$ of the time constant for the annealing. In addition, it is found in literature that the value of the $\beta$ constant decreases by $\sim  35$ \% in about two months after the end of the irradiation \cite{Kramberger_et_al_2007}. This is compared with an uncertainty of 23 \% in our measurement of the CCE variation. For all these reasons, we can neglect the effect of the damage annealing on the CCE variation measured for the SDD.

\section{Summary and conclusions}
\label{sec:conclusions}

We irradiated with a fluence of $7.9 \times 10^8 \; \mathrm{cm^{-2}}$ an SDD prototype (XDXL-1 production batch) during the assessment phase for LOFT as  candidate for the M3 launch opportunity. The average fraction of the beam intensity on the anodes was 88.8 \% (see the map in Fig. \ref{fig:beam_map}). The proton spectrum was centered at 11.2 MeV, with a FWHM of $\sim 6$ MeV.

After the irradiation, we measured a CCE decrease of $(0.65 \pm 0.15)$ \% between two positions at a distance of 30 mm along the drift channel. The measured value is in good agreement with the expected decrease of 0.8 \%, showing that the detectors can be safely employed in satellite-borne X-ray experiments without any degradation in CCE.

\section*{Acknowledgments}
LOFT is a project funded in Italy by ASI under contract ASI/INAF n. I/021/12/0 and in Switzerland through dedicated PRODEX contracts. The authors acknowledge the Italian INFN CSN5 for funding the Research and Development projects XDXL and REDSOX, and the INFN-FBK collaboration agreement MEMS2 under which the silicon drift detectors were produced. We gratefully acknowledge support of our measurements by the PIF team of PSI lead by W. Hajdas. The LOFT Consortium is grateful to the ESA Study Team for the professional and effective support to the assessment of the mission. This research has made use of NASA's Astrophysics Data System.







\end{document}